\documentclass[aps,prb,reprint,superscriptaddress,twocolumn]{revtex4-2}
\usepackage{amssymb,amsmath,bm,braket,txfonts,enumerate,mathtools,graphicx}
\usepackage[english]{babel}
\usepackage[justification=Justified, format=plain, font=small]{caption}
\usepackage[justification=Justified, format=plain, font=small]{subcaption}
\usepackage[bookmarks=true,colorlinks=true,citecolor=blue,urlcolor=blue,linkcolor=blue]{hyperref}
\usepackage{orcidlink}
\usepackage{pgfplots}
\usepackage{tikz}
\usepackage{color}
\usepackage{csquotes}
\usepackage[normalem]{ulem}
\usepgfplotslibrary{colorbrewer}
\usetikzlibrary{matrix} 
\pgfplotsset{compat=1.18}

\pgfplotsset{every axis/.append style={
                    label style={font=\small},
                    tick label style={font=\small},
                    tick pos=left
                    }}

\pgfplotsset{
  errorBars/.style={
    error bars/error bar style={
      line width=1.2pt,
    },
    error bars/y dir=both,
    error bars/y explicit,
    error bars/error mark=none,
  }
}
                    
\definecolor{7blue1}{HTML}{eff3ff}
\definecolor{7blue2}{HTML}{c6dbef}
\definecolor{7blue3}{HTML}{9ecae1}
\definecolor{7blue4}{HTML}{6baed6}
\definecolor{7blue5}{HTML}{4292c6}
\definecolor{7blue6}{HTML}{2171b5}
\definecolor{7blue7}{HTML}{084594}

\definecolor{5blue1}{HTML}{eff3ff}
\definecolor{5blue2}{HTML}{bdd7e7}
\definecolor{5blue3}{HTML}{6baed6}
\definecolor{5blue4}{HTML}{3182bd}
\definecolor{5blue5}{HTML}{08519c}

\definecolor{3blue1}{HTML}{6baed6}
\definecolor{3blue2}{HTML}{3182bd}
\definecolor{3blue3}{HTML}{08519c}

\definecolor{5orange1}{HTML}{feedde}
\definecolor{5orang2}{HTML}{fdbe85}
\definecolor{5orange3}{HTML}{fd8d3c}
\definecolor{5orange4}{HTML}{e6550d}
\definecolor{5orange5}{HTML}{a63603}

\definecolor{5red1}{HTML}{fee5d9}
\definecolor{5red2}{HTML}{fcae91}
\definecolor{5red3}{HTML}{fb6a4a}
\definecolor{5red4}{HTML}{de2d26}
\definecolor{5red5}{HTML}{a50f15}

\definecolor{4red1}{HTML}{fcbba1}
\definecolor{4red2}{HTML}{fc9272}
\definecolor{4red3}{HTML}{fb6a4a}
\definecolor{4red4}{HTML}{de2d26}

\definecolor{3red1}{HTML}{fb6a4a}
\definecolor{3red2}{HTML}{de2d26}
\definecolor{3red3}{HTML}{a50f15}

\definecolor{5green1}{HTML}{edf8e9}
\definecolor{5green2}{HTML}{bae4b3}
\definecolor{5green3}{HTML}{74c476}
\definecolor{5green4}{HTML}{31a354}
\definecolor{5green5}{HTML}{006d2c}

\definecolor{5gray1}{HTML}{f7f7f7}
\definecolor{5gray2}{HTML}{cccccc}
\definecolor{5gray3}{HTML}{969696}
\definecolor{5gray4}{HTML}{636363}
\definecolor{5gray5}{HTML}{252525}

\definecolor{5purple1}{HTML}{f2f0f7}
\definecolor{5purple2}{HTML}{cbc9e2}
\definecolor{5purple3}{HTML}{9e9ac8}
\definecolor{5purple4}{HTML}{756bb1}
\definecolor{5purple5}{HTML}{54278f}

\newcommand*{\figref}[1]{%
	\begingroup
	\hypersetup{
		linkcolor=blue,
		linkbordercolor=blue,
	}%
	\ref{#1}%
	\endgroup
}

\DeclareMathAlphabet{\mathbbb}{U}{bbold}{m}{n}
\DeclareMathOperator{\Tr}{Tr}
\let\Re\relax
\DeclareMathOperator{\Re}{Re}
\let\Im\relax
\DeclareMathOperator{\Im}{Im}
\DeclareMathOperator{\argmax}{argmax}
\DeclareMathOperator{\argmin}{argmin}
\DeclareMathOperator{\qf}{qf}
\newcommand{\id}{\mathbbb{1}}
\newcommand{\Stiefel}{\text{St}(n,p)}
\newcommand\thefontsize{The current font size is: \f@size pt}
\mathtoolsset{showonlyrefs}

\newcommand{\red}[1]{{#1}}

\begin{document}

\def\tum{{Technical University of Munich, TUM School of Natural Sciences, Physics Department, 85748 Garching, Germany}}
\def\mcqst{{Munich Center for Quantum Science and Technology (MCQST), Schellingstr. 4, 80799 M{\"u}nchen, Germany}}
\def\berkeley{{Department of Physics, University of California, Berkeley, CA 94720, USA}}
\def\berkeleymat{{Material Science Division, Lawrence Berkeley National Laboratory, Berkeley, CA 94720, USA}}
\newcommand{\TUM}{\affiliation{\tum}}
\newcommand{\MCQST}{\affiliation{\mcqst}}
\newcommand{\BERKELEY}{\affiliation{\berkeley}}
\newcommand{\BERKELEYMAT}{\affiliation{\berkeleymat}}
\title{Diagonal Isometric Form for Tensor Network States in Two Dimensions}
\author{Benjamin Sappler\orcidlink{0009-0002-4483-8094}}
\email{benjamin.sappler@tum.de} 
\TUM 
\MCQST
\author{Masataka Kawano\orcidlink{0000-0002-3477-2127}} \TUM \MCQST
\author{Michael P Zaletel\orcidlink{0000-0002-9297-7024}} \BERKELEY \BERKELEYMAT
\author{Frank Pollmann\orcidlink{0000-0003-0320-9304}} \TUM \MCQST
\date{\today}

\begin{abstract}
Isometric tensor network states (isoTNS) generalize the isometric form of the one-dimensional matrix product states to tensor networks in two and higher dimensions. Here, we introduce an alternative isometric form for isoTNS by incorporating auxiliary tensors to represent the orthogonality hypersurface. We implement the time evolving block decimation algorithm on this new isometric form and benchmark the method by computing ground states and the real time evolution of the transverse field Ising model in two dimensions on large square lattices of up to 1250 sites. Our results demonstrate that isoTNS can efficiently capture the entanglement structure of two-dimensional area law states. The short-time dynamics is also accurately reproduced even at the critical point. Our isoTNS formulation further allows for a natural extension to different lattice geometries, such as the honeycomb or kagome lattice.
\end{abstract}
\maketitle

\section{Introduction}
\label{sec:introduction}

Quantum many-body systems can exhibit many fascinating phenomena, including spin liquids with fractionalized excitations~\cite{tsui_two-dimensional_1982, laughlin_anomalous_1983, stormer_fractional_1999, han_fractionalized_2012}, topological phases of matter~\cite{wen_colloquium_2017}, and high-temperature superconductivity~\cite{bednorz_possible_1986,lee_doping_2006}. The numerical treatment of these systems is challenging due to the exponential growth of the Hilbert space dimension with system size. In one spatial dimension (1D) this challenge was addressed by the density matrix renormalization group (DMRG) algorithm~\cite{white_density_1992}, which can be understood as a variational method over the class of matrix product states (MPS)~\cite{dukelsky_equivalence_1998, schollwock_density-matrix_2011, orus_practical_2014}. The success of DMRG is due to the ability of MPS to efficiently capture the entanglement structure of gapped ground states~\cite{eisert_colloquium_2010, verstraete_matrix_2006}, which follows a characteristic area law~\cite{hastings_area_2007}. Additionally, every MPS can be brought into an isometric form~\cite{schollwock_density-matrix_2011, orus_practical_2014}, which greatly improves the computational cost of MPS algorithms. The natural generalization of MPS to higher dimensional systems is given in the form of projected entangled pair states (PEPS)~\cite{nishino_corner_1996, nishino_corner_1997, maeshima_vertical_2001, verstraete_renormalization_2004, levin_tensor_2007, jordan_classical_2008, jiang_accurate_2008, orus_practical_2014}. 
PEPS are able to efficiently represent area law states in two or more spatial dimensions~\cite{orus_practical_2014}. However, a general PEPS cannot be exactly brought into an isometric form similar to MPS due to the presence of closed loops. Consequently, algorithms on PEPS have computational costs scaling with high powers of the bond dimension $D$, e.g., ${\mathcal{O}(D^{10})}$ for a full time evolution update and ${\mathcal{O}(D^{12})}$ for DMRG~\cite{haegeman_unifying_2016, lubasch_algorithms_2014}. Additionally, DMRG on PEPS requires solving a generalized eigenvalue problem, which is numerically ill conditioned~\cite{lubasch_algorithms_2014}.\par
More recently, the new class of isometric tensor network states (isoTNS) has been introduced~\cite{zaletel_isometric_2020, haghshenas_conversion_2019, hyatt_dmrg_2020}, generalizing the isometric form of MPS to higher dimensions by enforcing isometry constraints on the tensors of a PEPS. This allows for the efficient computation of local expectation values and can reduce the computational cost of algorithms compared to PEPS. For example, the cost of real and imaginary time evolution is reduced from ${\mathcal{O}(D^{10})}$ to ${\mathcal{O}(D^7)}$~\cite{zaletel_isometric_2020}. While the generalization of DMRG to isoTNS still scales as ${\mathcal{O}(D^{12})}$, the generalized eigenvalue problem reduces to a standard eigenvalue problem due to the isometric form, which is much more stable numerically~\cite{lin_efficient_2022}. These advantages come at a cost, namely that the expressional power of isoTNS is lower than that of PEPS. 
\red{It was however found that isoTNS with finite bond dimension can exactly represent the ground state wave functions of string-net models~\cite{soejima_isometric_2020}. This suggests that long-range entanglement does not form an obstruction for isoTNS representations and that the ground states of gapped \red{2D} Hamiltonians with gappable edges can be efficiently represented as an isoTNS. Consequently, the isometric form of isoTNS is compatible with true 2D geometries}.
In Ref.~\cite{liu_simulating_2024}, topological phase transitions were studied with isoTNS, showing that the Ansatz is able to represent certain critical states with power-law correlations. isoTNS were also extended to fermionic systems~\cite{dai_fermionic_2024}, to 2D strips of infinite length~\cite{wu_two-dimensional_2023}, and to 3D cubic lattices~\cite{tepaske_three-dimensional_2021}. They have further been used to compute properties of 2D thermal states~\cite{kadow_isometric_2023}. Additionally, there have been works discussing the computational complexity of isoTNS contractions~\cite{malz_computational_2024} and relating isoTNS to quantum circuits~\cite{wei_sequential_2022, slattery_quantum_2021}.\par
While algorithms based on isoTNS have shown first promising results, there are still open questions. For example, the best way of defining the isometric form is not yet agreed upon. Different isometric forms could lead to reduced errors, more stable algorithms, and reduced computational cost. Recently, a new isometric form for isoTNS was introduced, where the direction of the isometries along columns alternates between pointing up and down~\cite{wu_alternating_2025}. This isometric form improves the ability of isoTNS to represent many-body states. In this work we propose yet another isometric form for isoTNS. The Ansatz differs from the original isoTNS introduced in Ref.~\cite{zaletel_isometric_2020} by rotating the lattice by ${45^\circ}$ and introducing auxiliary tensors with no physical degree of freedom.
\red{We find that our Ansatz yields ground-state energies for the transverse-field Ising model that are comparable to those obtained with the original isoTNS. The advantages of our Ansatz arise from the structural difference of the tensor network compared to the original isoTNS. Most isoTNS algorithms require repeatedly shifting a special column of tensors, the \textit{orthogonality hypersurface}, across the network. Because of the structure of our Ansatz, this shift becomes conceptionally simpler compared to the original isoTNS. Furthermore, while in our implementation the shift is still performed sequentially along one column of the network, one could also perform it in parallel at each tensor of the column. In that sense, moving the orthogonality hypersurface in our Ansatz can be understood as a local operation, compared to the more global shift in the original isoTNS. A further benefit of our Ansatz is the ease with which it generalizes to different lattice geometries; in this work, we demonstrate results on both square and honeycomb lattices.} \par
The paper is structured as follows. First, we give a review of MPS and isoTNS in 2D in Sec.~\ref{sec:isometric_tensor_network_states}. We discuss how the isometric form is able to speed up algorithms and review the Moses Move (MM)~ \cite{zaletel_isometric_2020}, which is a core part of isoTNS algorithms. In Sec.~\ref{sec:yb_isoTNS} we introduce the alternative isometric form and the Yang-Baxter (YB) move as the counterpart of the MM. We further implement an approximative algorithm for computing gradients during the disentangling step, giving similar results to the exact computation while being much faster. Additionally, we generalize the time evolving block decimation (TEBD) algorithm~\cite{vidal_efficient_2004}, which can be used for real and imaginary time evolution, to this new isometric form. In Sec.~\ref{sec:benchmarks} we benchmark the method by performing ground state search and real time evolution of the transverse field Ising model on large square and honeycomb lattices. Our results demonstrate that our Ansatz is able to efficiently capture the entanglement structure of 2D area-law states. We conclude with a discussion of our results in Sec.~\ref{sec:conclusion}.\par

\section{Isometric tensor network states}
\label{sec:isometric_tensor_network_states}
A general quantum many-body state can be written as
\begin{equation}
    \label{eq:quantum_many_body_wavefunction}
    \ket{\Psi} = \sum_{i_1=1}^{d_1} \sum_{i_2=1}^{d_2} \cdots \sum_{i_N=1}^{d_N} \Psi_{i_1,i_2,\dots,i_N} \ket{i_1}\otimes\ket{i_2}\otimes\cdots\otimes\ket{i_N},
\end{equation}
where $N$ is the number of subsystems (e.g., lattice sites or particles), ${\left\{\,\ket{i_1}\otimes\ket{i_2}\otimes\cdots\otimes\ket{i_N}\,\right\}}$ with ${i_n = 0,1,\dots,d_n}$ is a set of basis vectors of the full many-body Hilbert space
\begin{equation}
    \label{eq:many_body_hilbert_space}
    \mathcal{H} = \bigotimes_{n=1}^{N}\mathcal{H}_n,
\end{equation}
and ${d_n = \dim(\mathcal{H}_n)}$ is the dimension of the local Hilbert space ${\mathcal{H}_n}$ of subsystem $n$. To simplify the notation we assume that the dimension of all local Hilbert spaces is equal, ${d_n = d}$. The $d^N$ complex numbers ${\Psi_{i_1,i_2,\dots,i_N}}$ fully describe the quantum many-body state and one can think of ${\Psi\in\mathbb{C}^{d\times d\times\cdots\times d}}$ as a tensor of order $N$. However, due to the number of parameters scaling exponentially with system size, only very small system sizes are accessible computationally. \par
One can proceed by approximately decomposing $\Psi$ into a tensor network consisting of tensors of smaller order, with the total number of parameters depending polynomially on system size. If one is interested in computing ground state properties of local, gapped Hamiltonians, it is beneficial to choose a tensor network whose connectivity mimics the spatial structure of the physical system. This way, the tensor network is able to efficiently represent the area law entanglement structure present in the system~\cite{orus_practical_2014}. For 1D chains, the resulting tensor networks are MPS. For lattice systems of two or higher dimensions, PEPS are the natural generalization. We give a brief review of MPS in Sec.~\ref{sec:matrix_product_states} and of PEPS and isoTNS in 2D in Sec.~\ref{sec:MM_isoTNS}.
\subsection{Matrix product states in 1D}
\begin{figure}[t]
    \centering
    \begin{subfigure}[c]{\linewidth}
        \centering
        \caption{}
        \includegraphics[scale=1]{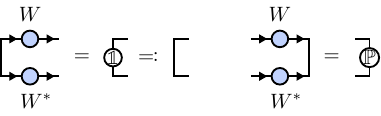}
        \label{fig:isometric_matrix}
    \end{subfigure}
    \begin{subfigure}[t]{0.4\linewidth}
        \centering
        \caption{}
        \includegraphics[scale=1]{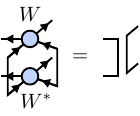}
        \label{fig:isometric_tensor}
    \end{subfigure}
    \begin{subfigure}[t]{0.59\linewidth}
        \centering
        \caption{}
        \includegraphics[scale=1]{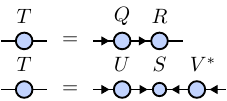}
        \label{fig:matrix_decompositions}
    \end{subfigure}
    \caption{Tensor network diagrams. (a) Isometry condition \eqref{eq:isometry_condition_general} for a tensor of order 2. (b) Isometry condition for a tensor of order 4. (c) QR decomposition and singular value decomposition.}
    \label{fig:basic_isometries}
\end{figure}
\label{sec:matrix_product_states}
By introducing an order-3 tensor ${T^{[n]}\in\mathbb{C}^{d\times\chi_{n-1}\times\chi_{n}}}$ for each subsystem $n$ and contracting the tensors in a chain, the wave function amplitudes in Eq.~\eqref{eq:quantum_many_body_wavefunction} can be written as an MPS
\begin{equation}
    \label{eq:MPS_definition}
    \Psi_{i_1,i_2,\dots,i_N} \coloneqq \sum_{\alpha_1=1}^{\chi_1} \sum_{\alpha_2=1}^{\chi_2} \dots \sum_{\alpha_{N-1}=1}^{\chi_{N-1}} T^{[1],i_1}_{1,\alpha_1} T^{[2],i_2}_{\alpha_1,\alpha_2} \dots T^{[N],i_{N}}_{\alpha_{N-1},1}
\end{equation}
with \textit{bond dimensions} $\chi_n$, where we have written the physical indices $i_n$ as superscripts, such that the sums are performed over subscripts only. The indices $\alpha_n$ are sometimes referred to as \textit{virtual indices}. Note that in this notation the bond dimensions at the two ends of the chain are ${\chi_0 = \chi_N = 1}$ and the tensors $T^{[1]}$ and $T^{[N]}$ can be interpreted as tensors of order $2$. \par
Any quantum state can be written as an MPS, where the maximum bond dimension ${\max(\{\chi_1,\dots,\chi_{N-1}\})}$ is exponential in system size. If the bond dimensions are truncated to a finite value ${\chi_n < \chi_\text{max}}$, the number of parameters of the MPS scales linearly in system size as ${\mathcal{O}(N\chi_\text{max}^2d)}$. Remarkably, all states with a 1D area law entanglement structure can be represented by an MPS with finite maximum bond dimension~\cite{orus_practical_2014, schollwock_density-matrix_2011}. \par
An important property of MPS is the existence of an \textit{isometric form} (also sometimes referred to as \textit{canonical form}), which speeds up most MPS algorithms. To bring an MPS into isometric form, one must turn the tensors $T^{[n]}$ into \textit{isometries}. An isometry is a matrix ${W\in\mathbb{C}^{m\times k}}$ with ${m\ge k}$ for which it holds
\begin{equation}
\label{eq:isometry_condition_general}
    W^\dagger W = \id,\quad WW^\dagger = \mathbb{P},
\end{equation}
with ${\mathbb{P} = \mathbb{P}^2}$ a projector to the smaller vector space. \par
It is beneficial to introduce a diagrammatic notation, in which tensors are drawn as shapes and indices as lines emerging from these shapes. Examples of this diagrammatic notation are shown in Fig.~\figref{fig:basic_isometries}. Tensor contractions are denoted by connecting the indices over which the contractions are performed. In tensor network diagrams, isometries can be denoted by decorating lines with arrows. In our convention incoming (outgoing) arrows correspond to the larger (smaller) dimension $m$ ($k$). The definition of isometries can easily be extended to tensors: an isometric tensor is a tensor which becomes an isometry by grouping together all legs with incoming and outgoing arrows, respectively. The isometry condition \eqref{eq:isometry_condition_general} is drawn in diagrammatic form for a matrix and a tensor of higher order in Fig.~\figref{fig:isometric_matrix} and Fig.~\figref{fig:isometric_tensor}, respectively.\par
Several decompositions can be used to decompose arbitrary matrices ${A\in\mathbb{C}^{n\times m}}$ into isometries. In this work, we use the QR decomposition and the singular value decomposition (SVD), which are depicted diagrammatically in Fig.~\figref{fig:matrix_decompositions}. \par
In the isometric form of MPS, an arbitrary site $n$ is chosen as the \textit{orthogonality center}. All other tensors are isometrized in such a way that all arrows point towards the orthogonality center. As an example, we draw a five-site MPS in isometric form in Fig.~\figref{fig:MPS_structure}. Any MPS can be exactly brought into isometric form using consecutive QR decompositions or SVDs. Further, the orthogonality center can be moved around easily and without error by contracting two neighboring tensors and splitting them again with a QR decomposition.\par
The isometric form improves the computational scaling and stability of most MPS algorithms. For example, consider the computation of the expectation value ${\bra{\Psi}\hat{O}_{ij}\ket{\Psi}}$ of a two-site operator $\hat{O}_{ij}$ acting on neighboring sites $i$ and $j$. This corresponds to the contraction of the tensor network that is created by sandwiching the operator between the bra and the ket. If the orthogonality center is at either site $i$ or site $j$, the isometry condition can be used to reduce the problem to the contraction of only five tensors; see Fig.~\figref{fig:MPS_expectation_value}.
\subsection{Isometric tensor network states in 2D}
\label{sec:MM_isoTNS}
\begin{figure}
    \centering
    \begin{subfigure}[t]{0.44\linewidth}
         \centering
         \caption{}
         \includegraphics[scale=1]{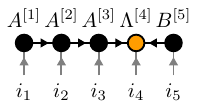}
         \label{fig:MPS_structure}
    \end{subfigure}
    \begin{subfigure}[t]{0.55\linewidth}
         \centering
         \caption{}
         \includegraphics[scale=1]{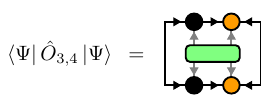}
         \label{fig:MPS_expectation_value}
    \end{subfigure}
    \begin{subfigure}[t]{0.44\linewidth}
         \centering
         \caption{}
         \includegraphics[scale=1]{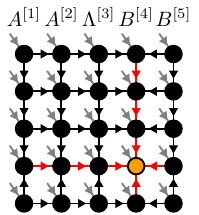}
         \label{fig:MM_isoTNS_structure}
    \end{subfigure}
    \begin{subfigure}[t]{0.55\linewidth}
        \centering
        \caption{}
        \raisebox{12.4pt}{%
            \includegraphics[scale=1]{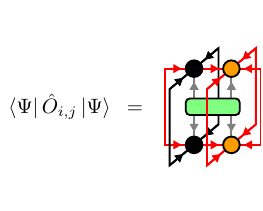}
        }%
        \label{fig:MM_isoTNS_expectation_value}
    \end{subfigure}
    \vspace{-8pt}
    \caption{Isometric tensor network states. (a) An MPS in isometric form representing a five-site chain is depicted as a tensor network diagram. (b) The expectation value of a local two-site operator reduces to a contraction of only five tensors due to the isometry condition \eqref{eq:isometry_condition_general}. (c) isoTNS on a 2D square lattice as introduced in Ref.~\cite{zaletel_isometric_2020}. The orthogonality hypersurface is drawn in red; the orthogonality center is drawn in orange. (d) Similar to MPS, local two-site expectation values at the orthogonality center reduce to a contraction of only five tensors due to the isometry condition.\vspace{-10pt}}
    \label{fig:MPS}
\end{figure}
The natural generalization of MPS to 2D is given by PEPS. A PEPS represents the local subsystem on each lattice site $j$ by the physical index $i_j$ of a tensor $T^{[j],i_j}$ and nearest-neighbor tensors are connected with virtual bonds according to the lattice structure. The maximum bond dimension of the virtual bonds is called $D$. The wave function amplitudes can be written as
\begin{equation}
    \Psi_{i_1,i_2,\dots,i_N} = \mathcal{C}\left(T^{[1],i_1},T^{[2],i_2},\dots,T^{[N],i_N}\right),
\end{equation}
where ${\mathcal{C}(\dots)}$ denotes the contraction of the full network over all virtual bonds. PEPS can efficiently represent states with an area-law entanglement structure in two and higher dimensions~\cite{orus_practical_2014}. Remarkably, PEPS can even represent certain states with correlations decaying polynomially with separation distance~\cite{verstraete_criticality_2006}, whereas MPS can only handle exponentially decaying correlations. Polynomially decaying correlations are characteristic for critical points. It is not generally possible to bring a PEPS into isometric form without error due to the presence of closed loops. Because of this, already the computation of local expectation values scales exponentially with system size and can only be computed approximately in practice, e.g., using the boundary MPS method~\cite{orus_practical_2014} or corner transfer matrices~\cite{baxter_series_1979}. Moreover, algorithms for ground state search and time evolution have computational costs scaling with high powers of the bond dimension. For example, the cost of a full TEBD update is dominated by the contraction of the effective environment, scaling as ${\mathcal{O}(D^{10})}$~\cite{lubasch_unifying_2014}. \par
isoTNS \cite{zaletel_isometric_2020,hyatt_dmrg_2020,haghshenas_conversion_2019} generalize the isometric form of MPS to higher dimensions by enforcing isometry constraints. A 2D isoTNS on the square lattice as defined in Ref.~\cite{zaletel_isometric_2020} is constructed by enforcing the isometry conditions shown in Fig.~\figref{fig:MM_isoTNS_structure}. The isometries are chosen in such a way that all arrows point towards a special row and column, called the orthogonality hypersurface of the isoTNS. Along the orthogonality hypersurface all arrows point towards the orthogonality center. The maximum bond dimension of bonds along the orthogonality hypersurface is increased to ${\chi = f\cdot D}$, where ${f\ge1}$ is a positive integer. In practice, this can produce better results at similar computational costs compared to increasing the maximum bond dimension for all bonds of the lattice~\cite{zaletel_isometric_2020,lin_efficient_2022}. Because of the isometry condition, one can think of the contractions of each of the four regions outside the orthogonality hypersurface as orthogonal boundary maps~\cite{lin_efficient_2022}. Local expectation values of operators acting in the vicinity of the orthogonality center can be computed efficiently because most contractions reduce to identity, similar to the computation of local expectation values in MPS; see Fig.~\figref{fig:MM_isoTNS_expectation_value}. \par
For implementing algorithms on isoTNS it is necessary to move the orthogonality center around the lattice. Moving the orthogonality center along the orthogonality hypersurface can be performed similar to MPS via QR decompositions. Moving the entire orthogonality hypersurface is a harder problem and can in general be only done approximately. In analogy to MPS, columns left of the orthogonality hypersurface are called $A^{[n]}$ and columns right of the orthogonality hypersurface are called $B^{[n]}$. Moving the orthogonality hypersurface $\Lambda^{[n]}$ one column to the right can be expressed as solving the problem
\begin{equation}
    \label{eq:MM_isoTNS_moving_OS_optimization_problem}
    \Lambda^{[n]}B^{[n+1]} \approx A^{[n]}\Lambda^{[n+1]},
\end{equation}
where the notation ${\Lambda^{[n]}B^{[n+1]}}$ denotes the contraction of the columns ${\Lambda^{[n]}}$ and $B^{[n+1]}$ along their connecting bonds. Instead of Eq.~\eqref{eq:MM_isoTNS_moving_OS_optimization_problem} one can solve the simpler auxiliary problem
\begin{equation}
\label{eq:MM_isoTNS_moving_os_auxiliary_optimization_problem}
    \Lambda^{[n]} = A^{[n]}\Lambda,
\end{equation}
\begin{figure}
    \centering
    \begin{subfigure}[b]{1.0\linewidth}
        \centering
        \caption{}
        \includegraphics[scale=1]{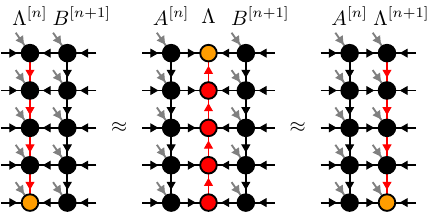}
        \label{fig:MM_isoTNS_moving_ortho_surface}
    \end{subfigure}
    \hfill
    \begin{subfigure}[b]{1.0\linewidth}
         \centering
         \caption{}
         \includegraphics[scale=1]{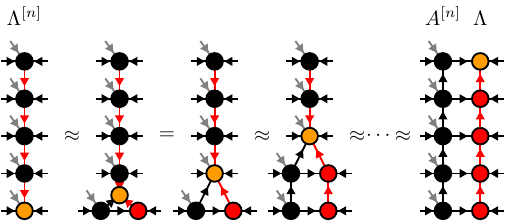}
         \label{fig:MM_isoTNS_MM_full}
    \end{subfigure}
    \hfill
    \begin{subfigure}[b]{1.0\linewidth}
         \centering
         \caption{}
         \includegraphics[scale=1]{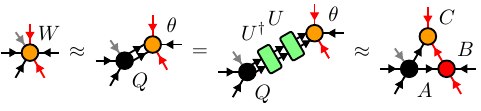}
         \label{fig:MM_isoTNS_MM_single}
    \end{subfigure}
    \vspace{-15pt}
    \caption{MM algorithm. (a) The orthogonality hypersurface can be moved in two steps as explained in the text. (b) The MM can be used to iteratively split the column $\Lambda^{[n]}$ into two columns $A^{[n]}$ and $\Lambda$ via repeated tripartite decompositions. (c) Tensor network diagram of a single tripartite decomposition as explained in the text.\vspace{-10pt}}
    \label{fig:MM_isoTNS_MM}
\end{figure}
where $\Lambda$ is a column of tensors without physical indices as shown in Fig.~\figref{fig:MM_isoTNS_moving_ortho_surface}. In a second step this column can then be absorbed into $B^{[n+1]}$ via the standard algorithm of applying a matrix product operator (MPO) to an MPS and compressing the resulting MPS to the maximal bond dimension~\cite{schollwock_density-matrix_2011} to form the new orthogonality hypersurface $\Lambda^{[n+1]}$~\cite{zaletel_isometric_2020}.
One can variationally solve problem Eq.~\eqref{eq:MM_isoTNS_moving_os_auxiliary_optimization_problem} by minimizing the distance ${\lVert\Lambda^{[n]}-A^{[n]}\Lambda\rVert}$, sweeping over all tensors of the columns $A^{[n]}$ and $\Lambda$ while performing local optimizations, which is known as an Evenbly-Vidal style variational optimization~\cite{evenbly_algorithms_2009, evenbly_algorithms_2014}.
It is however found in~\cite{zaletel_isometric_2020} that a single unzipping sweep, called the Moses Move (MM), provides a solution very close to the variational one while being far quicker. The MM can also be used as a good initialization for the variational algorithm.
We sketch the MM in Fig.~\figref{fig:MM_isoTNS_MM_full}. Starting from the bottom of the orthogonality hypersurface column, the tensors are split one after the other using tripartite decompositions.
A single tripartite decomposition of a tensor $W$ is shown in Fig.~\figref{fig:MM_isoTNS_MM_single}.
First, $W$ is split into two tensors $Q$ and $\theta$ via a truncated SVD ${W = U(SV) = Q\theta}$, where $Q$ is an isometry.
The bond connecting $Q$ and $\theta$ is then reshaped into two bonds of bond dimension ${\le D}$.
Next, it is important to note that the full contraction is invariant under the insertion of a unitary $U$ and its conjugate transpose, ${Q\theta = (QU^\dagger)(U\theta)}$, with ${A = QU^\dagger}$ still satisfying the isometry condition. This degree of freedom can be used to \textit{disentangle} the tensor $\theta$ along the vertical direction.
Accordingly, $U$ is chosen such that the truncation error or some entanglement measure is minimized for splits of $\theta$ along this direction. Choosing a good disentangling unitary is crucial for a successful tripartite decomposition.
It is found in Ref.~\cite{lin_efficient_2022} that an Evenbly-Vidal style variational minimization of the Rényi-2 entropy followed by a Riemannian optimization of the Rényi-${1/2}$ entropy gives the best results. It can be shown that the Rényi-$\alpha$ entropy in general bounds the truncation error for a fixed bond dimension if ${\alpha < 1}$~\cite{verstraete_matrix_2006}.
After contracting ${QU^\dagger}$ and ${U\theta}$, a truncated SVD is used to split ${U\theta}$ into tensors $B$ and $C$ as shown in the last step of Fig.~\figref{fig:MM_isoTNS_MM_single}, completing the tripartite decomposition. The computational cost of the MM scales with the maximum bond dimension $D$ as ${\mathcal{O}(D^7)}$~\cite{zaletel_isometric_2020,lin_efficient_2022}.\par
With the MM it is straightforward to generalize certain MPS algorithms to isoTNS. For example, real and imaginary time evolution can be performed via the TEBD$^2$ algorithm, which is a nested loop of TEBD performed on the orthogonality hypersurface (which can be thought of as MPS with an enlarged physical dimension), where in one iteration of TEBD$^2$ time evolution gates are applied to all nearest-neighbor pairs of tensors. In comparison to PEPS, the cost of TEBD$^2$ is reduced from ${\mathcal{O}(D^{10})}$ to ${\mathcal{O}(D^7)}$~\cite{zaletel_isometric_2020,lin_efficient_2022}.

\section{Alternative isometric form}
\label{sec:yb_isoTNS}

In the following, we introduce an alternative way of defining the isometric form for isometric tensor network states, which has some important differences to the isoTNS discussed in Sec.~\ref{sec:MM_isoTNS}. We start by introducing the new isometric form in Sec.~\ref{sec:YB_isoTNS_network_structure}. To move the orthogonality hypersurface, a similar algorithm to the MM is necessary, which we call the Yang-Baxter (YB) move and discuss in Sec.~\ref{sec:YB_isoTNS_YB_move}. In Section \ref{sec:YB_isoTNS_TEBD} we generalize the TEBD algorithm to this new isometric form.
\subsection{Network structure}
\label{sec:YB_isoTNS_network_structure}
\begin{figure}
    \centering
    \begin{subfigure}[c]{1.0\linewidth}
         \centering
         \caption{}
         \vspace{-10pt}
         \includegraphics[scale=1]{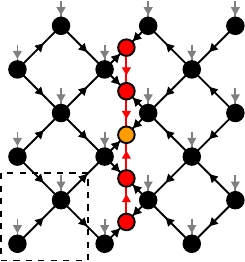}
         \label{fig:YB_isoTNS_structure}
    \end{subfigure}
    \centering
    \begin{subfigure}[c]{1.0\linewidth}
         \centering
         \caption{}
         \vspace{-10pt}
         \includegraphics[scale=1]{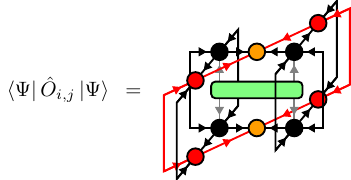}
         \label{fig:YB_isoTNS_expectation_value}
    \end{subfigure}
    \caption{(a) Tensor network diagram of an isoTNS in the alternative isometric form on a 2D square lattice. The orthogonality hypersurface is composed of auxiliary tensors with no physical degrees of freedom drawn in red. The orthogonality center is drawn in orange. The dashed lines denote a single unit cell of the lattice. (b) The computation of local expectation values at the orthogonality center reduces to a contraction of only a few tensors, similar to the original isoTNS [compare Fig.~\protect\figref{fig:MM_isoTNS_expectation_value}]. \vspace{-10pt}}
    \label{fig:YB_isoTNS_structure_and_expectation_value}
\end{figure}
The structure of an isoTNS in the alternative isometric form on a square lattice is shown in Fig.~\figref{fig:YB_isoTNS_structure}. It can be constructed in three steps. First, a square PEPS is rotated by 45$^\circ$. Next, the orthogonality hypersurface is constructed as a column of auxiliary tensors. The auxiliary tensors are connected in a line similar to an MPS and placed between two columns of PEPS tensors. Note that, in contrast to the original isoTNS, the tensors of the orthogonality hypersurface do not carry any physical degrees of freedom and only have virtual indices. Lastly, the isometry condition is enforced such that all arrows point towards the orthogonality hypersurface. Tensors left of the orthogonality hypersurface are thus brought into a left-isometric form and tensors right of the orthogonality hypersurface are brought into a right-isometric form, as shown in Fig.~\figref{fig:YB_isoTNS_structure}. The auxiliary tensors making up the orthogonality hypersurface are isometrized such that all arrows point towards a single auxiliary tensor---the orthogonality center. \par
In the following, we will denote the auxiliary tensors by $W_j$ and the tensors carrying physical degrees of freedom by $T_j$. The bonds connecting two $T$ tensors or a $T$ tensor and a $W$ tensor are truncated to a maximal bond dimension of $D$, while the maximal bond dimension between two $W$ tensors is denoted as $\chi$. Similar to the original isoTNS it is found that setting ${\chi = f\cdot D}$ with an integer ${f \ge1}$ produces good results in practice. The alternative isometric form again allows for the fast computation of expectation values of local operators; see Fig.~\figref{fig:YB_isoTNS_expectation_value}.
\subsection{Yang-Baxter move}
\label{sec:YB_isoTNS_YB_move}
\begin{figure}[b]
    \vspace{-10pt}
    \centering
    \begin{subfigure}[b]{1.0\linewidth}
        \centering
        \caption{}
        \includegraphics[scale=1]{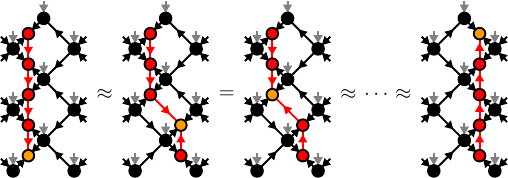}
    \label{fig:YB_isoTNS_shifting_ortho_surface}
    \end{subfigure}
    \centering
    \begin{subfigure}[b]{1.0\linewidth}
        \centering
        \vspace{-5pt}
        \caption{}
        \vspace{-7pt}
        \includegraphics[scale=1]{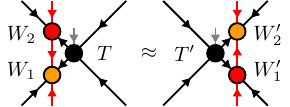}
        \label{fig:YB_move_single}
    \end{subfigure}
    \centering
    \begin{subfigure}[b]{1.0\linewidth}
        \centering
        \vspace{-5pt}
        \caption{}
        \vspace{-7pt}
        \includegraphics[scale=1]{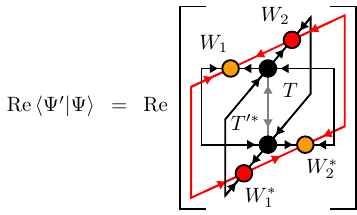}
        \label{fig:real_part_of_overlap_yb_move}
    \end{subfigure}
    \caption{(a) In the alternative isometric form, the orthogonality hypersurface can be moved one column to the right by iteratively applying YB moves. (b) Tensor network diagram of a single YB move. (c) The isometry condition reduces the overlap of the state before and after a single YB move to a contraction of six tensors.}
    \label{fig:YB_move_full_and_single}
\end{figure}
As with the original isoTNS, the orthogonality center can be easily and exactly moved along the orthogonality hypersurface using QR decompositions, but shifting the orthogonality hypersurface to the left and right is a harder problem. In analogy to the MM we look for a procedure to iteratively shift the orthogonality hypersurface through one column of $T$ tensors as shown in Fig.~\figref{fig:YB_isoTNS_shifting_ortho_surface}. A single iteration of this process is shown in Fig.~\figref{fig:YB_move_single}. The two tensors $W_1$ and $W_2$, which are part of the orthogonality hypersurface, are ``pulled through`` the tensor $T$, resulting in updated tensors $T^\prime$, $W_1^\prime$, and $W_2^\prime$. To keep the isometric structure of the network, $T^\prime$ and $W_1^\prime$ must be isometries and (if the wave function is normalized) $W_2^\prime$ must be a tensor of norm one. Due to the visual similarity to the Yang-Baxter equation we call this procedure the Yang-Baxter move. Accordingly, to distinguish the alternative isometric form from the original isoTNS, we will in the following refer to the original isoTNS as MM-isoTNS, whereas we will call isoTNS in the new isometric form YB-isoTNS. \par
We denote the state represented by the YB-isoTNS before the YB move by ${\ket{\Psi} = \ket{\Psi(W_1,W_2,T)}}$ and the state after the YB move by ${\ket{\Psi^\prime} = \ket{\Psi^\prime(W_1^\prime,W_2^\prime,T^\prime)}}$. One can think of the YB move as an optimization problem
\begin{equation}
    \label{eq:YB_move_optimization_problem}
    \left(T^\prime_\text{opt},W_{1,\text{opt}}^\prime,W_{2,\text{opt}}^\prime\right) = \underset{T^\prime,W_1^\prime,W_2^\prime}{\argmin}\left\lVert \ket{\Psi} - \ket{\Psi^\prime}\right\rVert
\end{equation}
under the constraints ${T^{\prime\dagger}T^\prime = \id}$, ${W_1^{\prime\dagger}W_1^\prime = \id}$, and ${\lVert W_2^\prime\rVert_\text{F} = 1}$ with the Frobenius norm ${\lVert\cdot\rVert_\text{F}}$. The error of the YB move can be rewritten as
\begin{equation}
	\label{eq:YB_isoTNS_YB_move_rewritten_error}
	\begin{split}
		\left\lVert \ket{\Psi} - \ket{\Psi^\prime} \right\rVert =& \sqrt{\braket{\Psi|\Psi} + \braket{\Psi^\prime|\Psi^\prime} - 2\Re\braket{\Psi^\prime|\Psi}} \\
		=& \sqrt{2 - 2\Re\braket{\Psi^\prime|\Psi}},
	\end{split}
\end{equation}
where in the second step we used the fact that the wave function is normalized to one: ${\braket{\Psi|\Psi} = \braket{\Psi^\prime|\Psi^\prime} = 1}$. It follows that the optimization problem of minimizing the error becomes the problem of maximizing the overlap
\begin{equation}
	\label{eq:YB_isoTNS_YB_move_alternative_formulation}
	(T_\text{opt}^\prime, W_{1,\text{opt}}^\prime, W_{2,\text{opt}}^\prime) = \underset{T,W_1^\prime,W_2^\prime}{\argmax}\Re\braket{\Psi^\prime|\Psi}
\end{equation}
under the previously defined constraints. Because the only tensors that are changed by the YB move are $W_1$, $W_2$, and $T$ and the three tensors make up a subregion of the full network with only incoming arrows, we can use the isometry condition to reduce the computation of the overlap ${\braket{\Psi^\prime|\Psi}}$ to a contraction of only six tensors as shown in Fig.~\figref{fig:real_part_of_overlap_yb_move}.\par
\begin{figure}[b]
    \centering
    \begin{subfigure}[b]{1.0\linewidth}
         \centering
         \vspace{-13pt}
         \caption{}
         \includegraphics[scale=1]{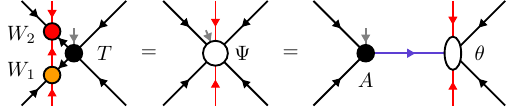}
         \label{fig:YB_move_tripartite_decomposition_a}
    \end{subfigure}
    \hfill
    \begin{subfigure}[b]{1.0\linewidth}
         \centering
         \caption{}
         \vspace{-7pt}
         \includegraphics[scale=1]{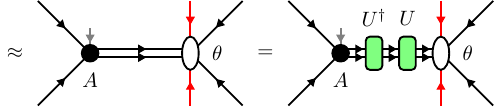}
         \label{fig:YB_move_tripartite_decomposition_b}
    \end{subfigure}
    \hfill
    \begin{subfigure}[b]{1.0\linewidth}
         \centering
         \caption{}
         \vspace{-7pt}
         \includegraphics[scale=1]{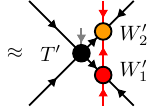}
         \label{fig:YB_move_tripartite_decomposition_c}
    \end{subfigure}
    \hfill
    \vspace{-16pt}
    \caption{Tensor network diagram of the tripartite decomposition algorithm that can be used for performing the YB move. The algorithm is explained in detail in the text.}
    \label{fig:yb_move_single_tripartite_decomposition}
\end{figure}
\begin{figure}[t]
    \centering
    \begin{subfigure}[b]{1.0\linewidth}
        \centering
        \caption{}
        \vspace{-10pt}
        \includegraphics[scale=1]{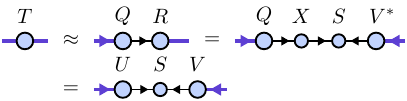}
        \label{fig:approximate_svd_full}
    \end{subfigure}
    \hfill
    \centering
    \begin{subfigure}[b]{1.0\linewidth}
        \centering
        \caption{}
        \vspace{-15pt}
        \includegraphics[scale=1]{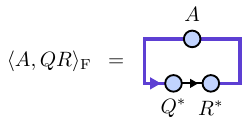}
        \label{fig:approximate_qr_overlap}
    \end{subfigure}
    \vspace{-5pt}
    \caption{(a) Approximate truncated SVD based on Ref.~\cite{unfried_fast_2023}. Bonds of bond dimension $\chi_1$ and $\chi_2$ are drawn in purple; bonds of bond dimension $\chi_\text{trunc}$ are drawn in black. (b) Tensor network diagram of the overlap ${\langle A,QR\rangle_\text{F}}$ that can be iteratively maximized to compute the approximate QR decomposition which is the first step of the approximate truncated SVD.\vspace{-15pt}}
    \label{fig:approximate_svd}
\end{figure}
To solve the optimization problem Eq.~\eqref{eq:YB_isoTNS_YB_move_alternative_formulation}, we implemented two explicit algorithms. The first algorithm is an Evenbly-Vidal style variational optimization method with iterative local updates. The second algorithm is a tripartite decomposition algorithm with disentangling, similar to the tripartite decomposition used in the MM. In practice we find that the latter performs much better. In the following we will thus only discuss the tripartite decomposition YB move; for more details on the first algorithm see Appendix \ref{app:optimization_of_isometric_tensor_networks}. \par
The tripartite decomposition YB move algorithm is similar to the tripartite decomposition used in the MM and is sketched in Fig.~\figref{fig:yb_move_single_tripartite_decomposition}. We start by contracting the tensors $T$, $W_1$, and $W_2$ into a single tensor $\Psi$. This tensor is then split from left to right via a truncated SVD ${\Psi = ASV^\dagger = A(SV^\dagger) \eqqcolon A\theta}$; see Fig.~\figref{fig:YB_move_tripartite_decomposition_a}. Here, the bond connecting $A$ and $\theta$ has a bond dimension of ${\min(dD^2, \chi^2D^2)}$. The bond dimension is truncated to $D^2$. Next, we split the index of the bond connecting $A$ and $\theta$ into two indices of dimension $D$ each. There exists a degree of freedom on the bonds connecting $A$ and $\theta$: a unitary $U$ and its adjoint can be inserted without changing the result of the contraction $A\theta = {AU^\dagger U\theta = (AU^\dagger)(U\theta) \eqqcolon T^\prime \tilde{\theta}}$; see Fig.~\figref{fig:YB_move_tripartite_decomposition_b}. The unitary $U$ can be chosen to minimize the truncation error of the next step by disentangling the tensor $\theta$. Similarly to MM-isoTNS~\cite{zaletel_isometric_2020, lin_efficient_2022} we find that a cheap Rényi-2 disentangling using an Evenbly-Vidal style variational algorithm followed by a Rényi-0.5 disentangling using Riemannian optimization gives the best results in practice. For more details on Evenbly-Vidal style and Riemannian optimization of isometric tensor networks see Appendix \ref{app:optimization_of_isometric_tensor_networks}. In the last step, Fig.~\figref{fig:YB_move_tripartite_decomposition_c}, the tensor $\tilde{\theta}$ is vertically split into $W_1^\prime$ and $W_2^\prime$ using a truncated SVD. Here, the bond dimension is truncated to $\chi$. We end up with the three tensors $T^\prime$, $W_1^\prime$, and $W_2^\prime$, completing the YB move. \par 
\red{We note that, because of the $45^\circ$ rotated network structure, the physical tensors involved in shifting the orthogonality surface by one site are not directly connected with each other in the tensor network; see Fig.~\ref{fig:YB_isoTNS_shifting_ortho_surface}. Because of this, one can think of the YB move as an entirely local operation, in contrast to the inherently iterative nature of the unzipping MM. A similar construction to the YB-isoTNS could also be done by inserting the orthogonality hypersurface as a column of auxiliary tensors without performing a 45$^\circ$ rotation of the square lattice. However, as long as the orthogonality hypersurface is not diagonal with respect to the lattice structure, the shift can only be performed iteratively and this locality is lost.}\par
The computational complexity of the YB move using the tripartite decomposition scales as ${\mathcal{O}(D^9)}$. However, this can be brought down to ${\mathcal{O}(D^8)}$ by using an approximation of the truncated SVD based on Ref.~\cite{unfried_fast_2023}. Assume we would like to perform the truncated SVD ${A \approx USV}$ with ${A\in\mathbb{C}^{\chi_1\times\chi_2}}$, truncating to a bond dimension ${\chi_\text{trunc} < \chi_1,\chi_2}$. The truncated SVD can be computed by performing a full SVD and keeping only the rows and columns corresponding to the ${\chi_\text{trunc}}$ largest singular values, scaling as ${\mathcal{O}(\chi_1\chi_2\min(\chi_1,\chi_2))}$. Alternatively, an approximate iterative procedure can be used, which we sketch in Fig.~\figref{fig:approximate_svd_full}. First, an approximate QR decomposition ${A = QR}$ with ${Q\in\mathbb{C}^{\chi_1\times\chi_\text{trunc}}}$, ${R\in\mathbb{C}^{\chi_\text{trunc}\times\chi_2}}$ is performed. For this, we first initialize the matrices $Q$ and $R$, e.g., using random values. We then iteratively maximize the overlap ${\langle A, QR\rangle_\text{F}}$ [see Fig.~\figref{fig:approximate_qr_overlap}], alternatingly optimizing the tensors $Q$ and $R$ at a computational complexity of ${\mathcal{O}(\chi_1\chi_2\chi_\text{trunc})}$ per iteration. After convergence, a SVD ${R = XSV}$ is performed at a cost of ${\mathcal{O}(\chi_2\chi_\text{trunc}^2)}$. Setting ${U = QX}$ concludes the approximate SVD at an overall cost of ${\mathcal{O}(N_\text{QR}\chi_1\chi_2\chi_\text{trunc})}$, where ${N_\text{QR}}$ denotes the maximum number of iterations for the approximate QR decomposition. \par
The approximate truncated SVD decomposition can also be used to approximate gradients used in the Riemannian optimization during the disentangling process, again bringing down the computational complexity from ${\mathcal{O}(D^9)}$ to ${\mathcal{O}(D^8)}$. This is particularly useful here because during the iterative Riemannian optimization algorithm, the current iterate for the disentanglement unitary changes only very little each iteration. This means that the result of the approximate QR decomposition of the previous iteration can be used as initialization for the current iteration, leading to a very fast convergence (in practice ${N_\text{QR}\le5}$ iterations) of the approximate QR decomposition.
\vspace{-15pt}
\subsection{TEBD\texorpdfstring{\textsuperscript{2}}{2}}
\label{sec:YB_isoTNS_TEBD}
\begin{figure}[t]
    \centering
    \includegraphics[scale=1]{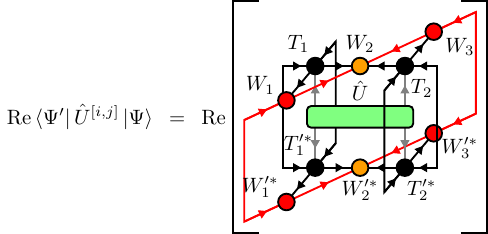}
    \caption{Overlap of the state before and after applying a local TEBD gate ${\hat{U}^{[i, j]}}$ is reduced to the contraction of only a few tensors due to the isometry condition.\vspace{-10pt}}
    \label{fig:real_part_of_overlap_tebd_update}
\end{figure}
We will now discuss the generalization of the time evolving block decimation (TEBD) algorithm~\cite{vidal_efficient_2004} to YB-isoTNS. The TEBD algorithm can be used for both real and imaginary time evolution. Analogously to MPS we start with a Suzuki-Trotter decomposition, approximating the time evolution operator ${\hat{U}\left(\Delta t\right) = e^{-i\Delta t \hat{H}}}$ by a product of bond operators ${\hat{U}^{[i, j]}\left(\Delta t\right) = e^{-i\hat{h}^{[i,j]}\Delta t}}$ acting only on neighboring sites $i$ and $j$ of the bond ${[i, j]}$. These bond operators then must be applied to the state in the correct order, while keeping the YB-isoTNS structure intact. A full update is performed by applying all bond operators, which approximately evolves the state by time $\Delta t$. We will first discuss the process of applying a single bond operator ${\hat{U}^{[i, j]}\left(\Delta t\right)}$ to the YB-isoTNS. Let us assume that the orthogonality center is positioned between the two sites on which the bond operator ${\hat{U}^{[i, j]}\left(\Delta t\right)}$ acts. The five tensors around the orthogonality center then make up a sub network with only incoming arrows; see Fig.~\figref{fig:real_part_of_overlap_tebd_update}. We call these five tensors $T_i$, $T_j$, $W_1$, $W_2$, and $W_3$. The local TEBD update can then be formulated as the following problem: find tensors $T_i^\prime$, $T_j^\prime$, $W_1^\prime$, $W_2^\prime$, and $W_3^\prime$ satisfying the isometry constraints and minimizing the error
\begin{equation}
	\label{eq:isoDTNS_TEBD_local_update_error}
	\varepsilon_\text{trunc} = \left\lVert \hat{U}^{[i, j]}(\Delta t)\ket{\Psi} - \ket{\Psi^\prime}\right\rVert.
\end{equation} 
Similar to the YB move, we can rewrite this as the problem of maximizing the overlap
\begin{equation}
    \label{eq:YB_isoTNS_TEBD_maximizing_overlap}
    \begin{split}
        (T_{i,\text{opt}}^\prime, T_{j,\text{opt}}^\prime, &W_{1,\text{opt}}^\prime, W_{2,\text{opt}}^\prime, W_{3,\text{opt}}^\prime)\\
        &= \underset{T_i^\prime, T_j^\prime, W_1^\prime, W_2^\prime, W_3^\prime}{\argmax}\Re\bra{\Psi^\prime}\hat{U}^{[i,j]}(\Delta t)\ket{\Psi}
    \end{split}
\end{equation}
under the constraints ${T_i^{\prime\dagger}T_i^\prime = \id}$, ${T_j^{\prime\dagger}T_j^\prime = \id}$, ${W_1^{\prime\dagger}W_1^\prime = \id}$, ${W_3^{\prime\dagger}W_3^\prime = \id}$, and ${\lVert W_2^\prime\rVert_\text{F} = 1}$. Using the isometry condition, the overlap ${\bra{\Psi^\prime}\hat{U}^{[i, j]}(\Delta t)\ket{\Psi}}$ can be computed by contracting the tensor network drawn in Fig.~\figref{fig:real_part_of_overlap_tebd_update}. \par
For solving the constrained optimization problem  Eq.~\eqref{eq:YB_isoTNS_TEBD_maximizing_overlap} we use an Evenbly-Vidal style iterative optimization algorithm.
We optimize one tensor at a time while keeping all other tensors fixed.
This procedure is repeated, sweeping over all five tensors until convergence is achieved. For more details on this optimization method, see Appendix \ref{app:optimization_of_isometric_tensor_networks}. 
Since the time step ${\Delta t}$ is chosen to be small, the bond operator is close to identity, ${\hat{U}^{[i,j]}(\Delta t)\approx\id}$. Thus a good initialization for the tensors of the updated wave function ${\ket{\Psi^\prime}}$ are simply the tensors of the old wave function ${\ket{\Psi}}$. The computational complexity of applying a local bond operator to a YB-isoTNS with the discussed algorithm scales as ${\mathcal{O}(D^6)}$. In practice it is observed that the algorithm often converges after only a few iterations.\par
\onecolumngrid

\begin{figure*}[b]
    \vspace{-5pt}
    \centering
    \includegraphics[scale=1]{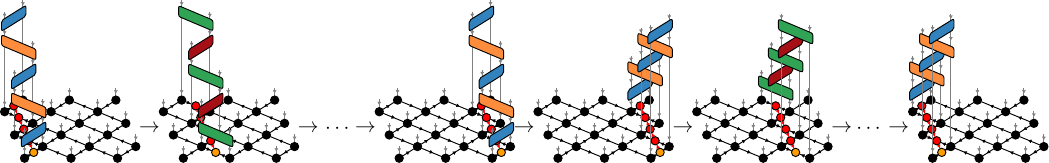}
    \caption{Full second order TEBD update of time step ${\Delta t}$ is performed by applying nearest-neighbor time evolution gates ${\hat{U}^{[i,j]}(\Delta t/2)}$ along the orthogonality hypersurface, moving from the leftmost to the rightmost position and back. When moving back, the gates are applied in reverse order.\vspace{-5pt}}
    \label{fig:yb_tebd2}
\end{figure*}
\clearpage
\begin{figure*}[t]
    \centering
    \begin{subfigure}[t]{0.82\textwidth}
        \caption{}
		\centering
        \vspace{-5pt}
        \hspace{-20pt}
		\includegraphics[scale=1]{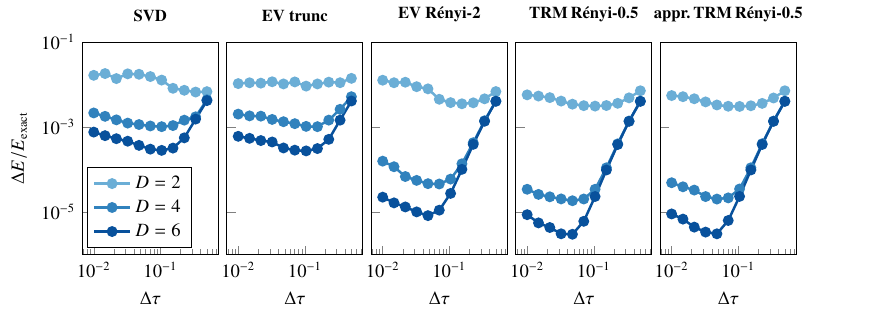}
	    \label{fig:tfi_gs_energy_vs_dtau_different_methods}
    \end{subfigure}
    \hspace{-20pt}
    \begin{subfigure}[t]{0.17\textwidth}
        \caption{}
        \centering
        \vspace{-5pt}
        \hspace{-50pt}
        \begin{minipage}{\textwidth}
    		\centering
		      \includegraphics[scale=1]{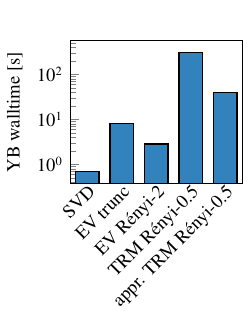}
        \end{minipage}
        \label{fig:tfi_gs_energy_vs_dtau_different_methods_walltimes}
    \end{subfigure}
    \vspace{-5pt}
    \caption{(a) In this figure we perform ground state search of the TFI model on a ${4\times4}$ diagonal square lattice ($32$ spins) at a transverse field of ${g = 3.5}$ using the imaginary time TEBD$^2$ YB-isoTNS algorithm. The relative energy error is plotted against the time step size ${\Delta \tau}$. The five plots correspond to five different algorithms used for performing the YB move; details are given in the text. For all iterative algorithms, we used a maximum of ${N_\text{iter,YB} = 100}$ iterations. For all plots, the bond dimension along the orthogonality surface was set to ${\chi=6\cdot D}$ and the number of iterations for local TEBD updates was set to ${N_\text{iter,TEBD} = 100}$. (b) We compare the duration that the different algorithms spend on the YB move during a full TEBD update of time ${\Delta\tau}$. We compute this accumulated walltime by running TEBD for ${N_\text{TEBD} = 10}$ steps, measuring the time spent performing YB moves, and computing the average walltime per TEBD step.\vspace{-10pt}%
    }%
    \label{fig:tfi_gs_energy_vs_dtau_different_methods_and_walltimes}
\end{figure*}
\twocolumngrid
To apply a global TEBD update, we must apply local TEBD updates at all bonds. Because each YB move introduces an error ${\varepsilon_\text{YB}}$, it is beneficial to minimize the number of necessary YB moves. By moving the orthogonality hypersurface once from left to right and back it is possible to apply a second order global TEBD update with a Trotterization error of ${\mathcal{O}(\Delta t^3)}$; see Fig.~\figref{fig:yb_tebd2}.
\vspace{-10pt}
\section{Benchmarking}
\label{sec:benchmarks}
To benchmark the method, we consider the transverse field Ising (TFI) model
\begin{equation}
    \label{eq:TFI_Hamiltonian}
	\hat{H}_\text{TFI} = -J\sum_{\langle i,j\rangle} \hat{\sigma}^x_i \hat{\sigma}^x_j - g\sum_{i} \hat{\sigma}^z_i,
\end{equation}
where $\langle i,j\rangle$ denotes pairs of nearest-neighbor spin-1/2 particles and $\hat{\sigma}^x_i, \hat{\sigma}^z_i$ are Pauli matrices. On the square lattice the model exhibits a quantum phase transition at a critical transverse field of $g_\text{C} \approx 3.04438$~\cite{blote_cluster_2002}. \par
We first compute the ground state of the TFI model Eq.~\eqref{eq:TFI_Hamiltonian} using YB-isoTNS and TEBD with imaginary time $\Delta t = -i\Delta\tau$, $\Delta\tau\in\mathbb{R}$. As an initial state, we choose the product state $\ket{\Psi_i}=\ket{\uparrow}\otimes\cdots\otimes\ket{\uparrow}$ of all spins pointing up. To use the TEBD algorithm, one must choose a good step size $\Delta\tau$. The total error of a single TEBD step is a sum of three errors,
\begin{equation}
	\varepsilon_\text{TEBD} = \varepsilon_\text{Trotter} + \varepsilon_\text{trunc} + \varepsilon_{\text{YB}}.
\end{equation}
The Trotterization error $\varepsilon_\text{Trotter}$ comes from the Suzuki-Trotter decomposition, the truncation error $\varepsilon_\text{trunc}$ comes from the local application of the TEBD bond operators, and the YB error $\varepsilon_{\text{YB}}$ arises when shifting the orthogonality hypersurface. A smaller time step $\Delta t$ decreases both $\varepsilon_\text{Trotter}$ and $\varepsilon_\text{trunc}$, while the YB error $\varepsilon_{\text{YB}}$ is not directly affected. However, since for a smaller time step $\Delta t$ more TEBD iterations are necessary to approach the limit $t\rightarrow\infty$, YB errors add up and prevent the state from reaching the true ground state. Therefore, we expect, similar to~\cite{zaletel_isometric_2020, lin_efficient_2022}, that the YB error dominates for small time steps, while the trotterization and truncation errors dominate for larger time steps. The best results can be achieved when the time step $\Delta t$ is tuned such that $\varepsilon_\text{Trotter} + \varepsilon_\text{trunc} \approx \varepsilon_{\text{YB}}$. \par
In Fig.~\figref{fig:tfi_gs_energy_vs_dtau_different_methods} we plot the obtained relative ground state energy error against the time step $\Delta\tau$ on a $4\times4$ diagonal square lattice at a transverse field of $g=3.5$. The numerically exact reference energy was computed using an MPS-DMRG simulation using \textit{tenpy}~\cite{hauschild_efficient_2018}, snaking an MPS through the 2D lattice. We compare the following five different algorithms for performing the YB move.
\begin{enumerate}[(i)]
    \item \textbf{SVD}. Tripartite decomposition via two consecutive SVDs without disentangling;
    \item \textbf{EV trunc}. Direct Evenbly-Vidal style optimization of the overlap Eq.~\eqref{eq:YB_isoTNS_YB_move_alternative_formulation} (directly minimizing the error $\varepsilon_\text{YB}$).
    \item \textbf{EV Rényi-2}. Tripartite decomposition with Evenbly-Vidal style Rényi-2 disentangling.
    \item \textbf{TRM Rényi-0.5}. Same as \textbf{EV Rényi-2}, but with additional disentangling using the Riemannian Trust Region Method, a quasi-Newton optimization method (see Appendix \ref{app:Riemannian_optimization} for more details).
    \label{item:TRM_Renyi_0_5}
    \item \textbf{appr.\,TRM Rényi-0.5}. Same as \textbf{TRM Rényi-0.5}, but using approximate truncated SVD for computing the gradients.
    \label{item:TRM_Renyi_0_5_approx}
\end{enumerate}
In Fig.~\figref{fig:tfi_gs_energy_vs_dtau_different_methods_walltimes} we additionally plot the YB move walltimes of the different algorithms. We observe that disentangling plays a crucial role in minimizing the error of the YB move. (\ref{item:TRM_Renyi_0_5}) \textbf{TRM Rényi-0.5} gives the best results in practice, but has a large computational cost. Using the approximate gradient computation gives essentially the same results, but speeds up the algorithm by almost an order of magnitude. In all following benchmarks we will thus use (\ref{item:TRM_Renyi_0_5_approx}) \textbf{appr.\,TRM Rényi-0.5} for performing the YB move.\par
\begin{figure}[t]
\def\xmin{4.5}%
\def\xmax{25.5}%
\def\ymin{-3.652}%
\def\ymax{-3.608}%
\def\xminzoomed{19.75}%
\def\xmaxzoomed{25.25}%
\def\yminzoomed{-3.6496}%
\def\ymaxzoomed{-3.646}%
\def\insetLineOpacity{1.0}%
\def\insetLineWidth{0.5pt}
\begin{subfigure}[c]{1.0\linewidth}
    \caption{}
    \centering
    \vspace{-10pt}
    \hspace{70pt}
        \includegraphics[scale=1]{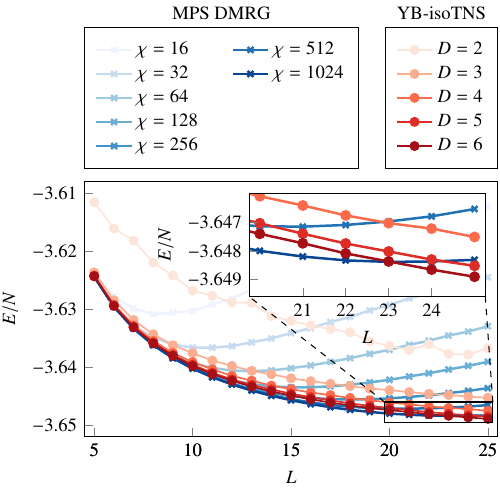}
        \vspace{-12pt}
        \label{fig:YB_isoTNS_gs_search_larger_systems}
\end{subfigure}
\begin{subfigure}[c]{1.0\linewidth}
    \caption{}
    \centering
    \hspace{0.8pt}
    \includegraphics[scale=1]{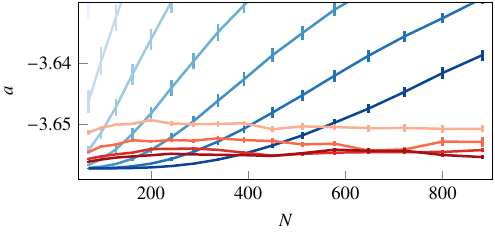}
    \label{fig:YB_isoTNS_gs_search_larger_systems_a_fit}
\end{subfigure}
\caption{(a) In this figure we plot the lowest energy densities found using imaginary time TEBD$^2$ on the TFI model at $g=3.5$ [corresponding to the minima in Fig.~\protect\figref{fig:tfi_gs_energy_vs_dtau_different_methods}] against linear system size $L$. The algorithm parameters are the same as in Fig.~\protect\figref{fig:tfi_gs_energy_vs_dtau_different_methods}, except that we here use only $N_\text{iters,YB}=10$ iterations for the approximate TRM disentangling during the YB move. The reference simulation shown in blue was performed using MPS DMRG. We note that for large systems $L\ge20$ the reference MPS simulations with $\chi\ge256$ are only converged up to a ground state energy density error of $\Delta (E/N)\le10^{-4}$ due to large computation times and memory costs, which does however not change the results in a qualitative way. (b) Fit parameter $a$ obtained by fitting the function ${E = a\cdot N+ b\cdot N_\text{boundary}+c}$ to the data from (a). We used a rolling window of nine data points around each $N$ for the fit. The $D=2$ data was excluded from this plot due to noise. The error bars are the square root of the variance.}
\end{figure}

In Fig.~\figref{fig:YB_isoTNS_gs_search_larger_systems} we plot the best obtained ground state energy density against the linear system size $L$. Since the diagonal square lattice has a unit cell containing two sites, the number of spins is given by $N=2L^2$. In the limit $L\rightarrow\infty$ we expect the energy density $E/N$ to approach a constant, whereas the energy density of small systems is dominated by finite size effects. We observe that the energy density computed using the reference MPS-DMRG simulation first decreases but then increases again when going to larger system sizes, which happens earlier for smaller bond dimensions. The reason for this is that the inherently 1D MPS fails to capture the 2D area law entanglement structure of the ground state. More specifically, certain pairs of sites that are close together in the 2D lattice are far apart in the MPS that is snaked through the lattice, resulting in long-range interactions. In contrast, the energy density computed using YB-isoTNS does not show this effect and approaches a constant energy density already for $D = 2$. We interpret this as the YB-isoTNS being able to correctly capture the 2D area law entanglement structure. For large systems of size $L=24$ and $L=25$, YB-isoTNS with $D = 5$ and $D = 6$ is able to find lower ground state energies than MPS DMRG with $\chi=1024$, which can be seen in the inset in Fig.~\figref{fig:YB_isoTNS_gs_search_larger_systems}.
This is possible even though the isoTNS has much less variational parameters: for $L=25$, an MPS with a maximum bond dimension of $\chi=1024$ needs to store $N_\text{params,MPS} \approx 2.6\times10^9$ complex numbers, while a YB-isoTNS with $D=5$, $\chi=30$ only consists of $N_\text{params,YB-isoTNS}\approx2.7\times10^6$ complex numbers, which is a difference of three orders of magnitude. Similar to PEPS we expect the advantage of isoTNS to become even more apparent for even larger systems or closer to the critical point, where capturing the area law entanglement structure is more important. \par
We now perform a more careful analysis of the data from Fig.~\figref{fig:YB_isoTNS_gs_search_larger_systems}. The ground state energy of a gapped model with open boundary conditions is expected to scale as
\begin{equation}
    \label{eq:energy_scaling_gapped_model}
    E = a\cdot N + b\cdot N_\text{boundary} + c + ...,
\end{equation}
where $N_\text{boundary} \propto L$ and the remaining corrections are exponentially small in system size. In Fig.~\figref{fig:YB_isoTNS_gs_search_larger_systems_a_fit} we plot the fit parameter $a$ obtained by fitting Eq.~\eqref{eq:energy_scaling_gapped_model} over a rolling window around a given $N$. We observe that the MPS data shows a plateau for small system sizes $N \lesssim 200$ and large bond dimensions $\chi\geq 512$, where the simulation is numerically exact. The parameter $a$ diverges for larger system sizes. In contrast, the isoTNS data plateaus for large system sizes even for small bond dimensions $D$. However, the numerically exact value for $a$ that can be extracted from the MPS data at $N \lesssim 200$ is not reached even at $D=6$. We conclude that, for small system sizes, MPS performs better due to the fact that one can choose much larger bond dimensions compared to isoTNS. However, MPS fails to describe the model for large system sizes while the isoTNS simulation correctly captures the area law entanglement structure. \red{We emphasize that the strong performance of MPS in Fig.~\figref{fig:YB_isoTNS_gs_search_larger_systems} relies on the fact that the correlation length of the TFI model decreases quickly when moving away from the critical point. We expect the advantages of isoTNS to become more evident closer to criticality or for models that exhibit higher entanglement.}\par
\begin{figure}
    \begin{subfigure}[t]{0.75\linewidth}
        \caption{}
		\centering
        \hspace{-45pt}
		\includegraphics[scale=1]{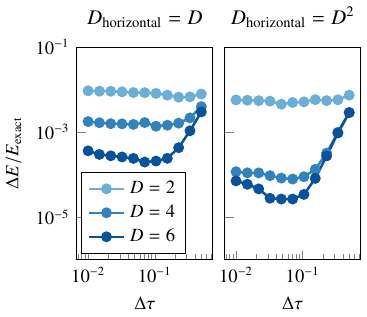}
        \label{fig:tfi_gs_energy_vs_dtau_honeycomb}%
        \end{subfigure}%
        \hspace{-15pt}
    \begin{subfigure}[t]{0.24\linewidth}%
        \caption{}
        \vspace{15pt}
		\includegraphics[scale=0.6]{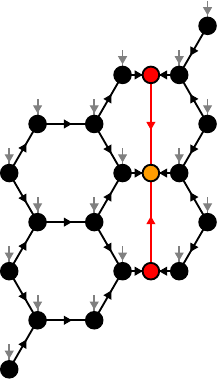}
        \label{fig:YB_isoTNS_honeycomb}
    \end{subfigure}
    \hfill
    \caption{(a) Imaginary time TEBD results using YB-isoTNS on the honeycomb lattice. The simulation for the left and right plot was computed using ${D_\text{horizontal} = D}$ and ${D_\text{horizontal} = D^2}$, respectively. The bond dimension along the orthogonality hypersurface was chosen as ${\chi = 6\cdot D}$. For the YB move we used the approximate Rényi-0.5 disentangler with a maximum of ${N_\text{iter,YB} = 100}$ iterations per YB move. The model is the TFI model on a ${4\times4}$ honeycomb lattice at a transverse field of ${g = 3.5}$. (b) YB-isoTNS network structure on a ${3\times3}$ honeycomb lattice.}
    \label{fig:tfi_gs_energy_vs_dtau_honeycomb_and_YB_isoTNS_honeycomb}
\end{figure}
To show that the algorithm can be generalized to different lattice structures we perform a ground state search of the TFI model on a honeycomb lattice in Fig.~\figref{fig:tfi_gs_energy_vs_dtau_honeycomb}. The resulting isoTNS has two different kinds of virtual bonds, horizontal and diagonal bonds, and is sketched in Fig.~\figref{fig:YB_isoTNS_honeycomb}. Since a single horizontal bond splits into two diagonal bonds it is intuitive to choose a larger bond dimension at the horizontal bonds, e.g., $D_\text{horizontal} = D^2$. \par
\begin{figure}
	\centering
	\begin{minipage}{1.0\linewidth}
		\includegraphics[scale=1]{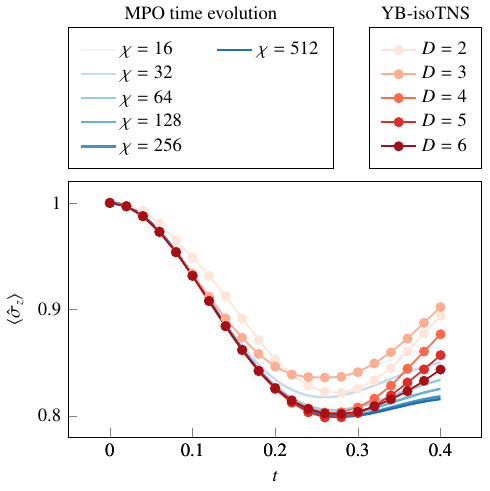}
	\end{minipage}
	\caption{In this figure we show the time evolution of the ${\langle\hat{\sigma}_z\rangle}$ expectation value of a spin in the middle of an ${8\times8}$ diagonal square lattice (at position $x=y=4$, $m=0$, where $m$ is the index into the unit cell), containing in total ${N = 128}$ spins. We use the TFI model at critical field $g_C$. We compute the time evolution once with the MPO time evolution algorithm~\cite{zaletel_time-evolving_2015} on a MPS and once with TEBD on a YB-isoTNS. We observe good agreement up to times of ${t \approx 0.25}$, where the two simulations diverge. For the isoTNS simulation we use a time step of $\Delta t = 0.02$, $N_\text{iter,TEBD} = 100$ iterations for local TEBD updates and $N_\text{iter,YB} = 100$ iterations of the (\ref{item:TRM_Renyi_0_5_approx}) \textbf{appr.\,TRM Rényi-0.5} disentangler for the YB move. The bond dimension along the orthogonality hypersurface is set to $\chi = 6\cdot D$. For the MPS reference simulation we use the second order MPO time evolution algorithm with a time step of $\Delta t = 0.001$.}
    \label{fig:YB_isoTNS_time_evolution_g_critical}
\end{figure}

 
We lastly study the capabilities of YB-isoTNS to perform real-time evolution. For this we perform a global quench by initializing the YB-isoTNS in the all-up state ${\ket{\Psi} = \ket{\uparrow}\otimes\cdots\otimes\ket{\uparrow}}$, which we then evolve in time using TEBD. We work on the $8\times8$ diagonal square lattice and evolve with the Hamiltonian of the TFI model at critical transverse field $g \approx 3.04438$. At the critical field the entanglement grows very quickly, making this a hard problem. We choose a step size of $\Delta t = 0.02$ and evolve up to time $t = 0.4$, requiring 20 TEBD iterations. For a comparison we use the second order MPO time evolution algorithm~\cite{zaletel_time-evolving_2015} defined on MPS, which is able to perform time evolution in the presence of long range interactions and is implemented in tenpy~\cite{hauschild_efficient_2018}. Due to the long-range couplings present in the effective Hamiltonian for the MPS a very small time step is necessary, leading to long computation times. We choose a time step of $\Delta t = 0.001$, which is computationally feasible but still not fully converged. We plot the expectation value ${\langle\hat{\sigma}^z\rangle}$ of a spin in the middle of the lattice against the time in Fig.~\figref{fig:YB_isoTNS_time_evolution_g_critical}. We observe that the YB-isoTNS results are in good agreement with the reference simulation up to a time of $t\approx0.25$, at which the two simulations diverge. This happens at earlier times for smaller bond dimensions $D$. We expect the reason for this to be the fast accumulation of errors during YB moves. We conclude that the short-time behavior of the system is correctly captured by the YB-isoTNS.

\section{Conclusion}
\label{sec:conclusion}
In this work we introduced and benchmarked a new Ansatz for defining the isometric form of isometric tensor network states in 2D. In the new Ansatz the orthogonality hypersurface is made up of auxiliary tensors without physical degrees of freedom. The YB move was introduced as an algorithm for moving the orthogonality hypersurface. In comparison to the MM used in previous works on isoTNS, the YB move does not need an additional step for MPO-MPS multiplication and compression. It was found, similar to the MM~\cite{zaletel_isometric_2020, lin_efficient_2022}, that it is crucial to perform a disentangling step during the YB move. The best results were achieved with a Riemannian optimization of the Rényi entropy. To speed up the YB move we implemented an approximate version of this algorithm, bringing down the computational complexity from ${\mathcal{O}(D^9)}$ to ${\mathcal{O}(D^8)}$. We finally implemented a TEBD$^2$ algorithm that can be used to perform real and imaginary time evolution on YB-isoTNS. \par
The algorithm was then benchmarked by performing ground state search and a global quench of the transverse field Ising model. 
It was found that YB-isoTNS are able to produce similar results to MM-isoTNS. 
The algorithms performing best in the ground state search were found to be the Riemannian optimization algorithms minimizing the Rényi entropy with ${\alpha = 1/2}$.
In a large scale simulation of up to ${N = 1250}$ spins, numerical evidence was found that YB-isoTNS are able to correctly capture the area law entanglement structure of the ground state. 
For the largest systems we looked at, YB-isoTNS was able to achieve lower ground state energies while needing to store $\approx10^3$ times less variational parameters compared to MPS. 
Additionally, the Ansatz was used to compute the ground state energy of the TFI model on a honeycomb lattice, showcasing that YB-isoTNS can be easily generalized to other lattice types. 
Computing a real time evolution after a global quench was found to be challenging for YB-isoTNS. The reason for this is the fast accumulation of YB move errors. 
Nevertheless, YB-isoTNS were found to be able to correctly capture short-time behavior of the 2D TFI model with ${N = 128}$ spins even at criticality. 
Additionally, a much larger time step $\Delta t$ can be used compared to the MPO time evolution algorithm on MPS, leading to shorter computation times. \par
As for the performance, the YB-isoTNS Ansatz is able to produce comparable results to MM-isoTNS. While the computational complexity of shifting the orthogonality hypersurface of YB-isoTNS is larger [${\mathcal{O}(D^8)}$ compared to ${\mathcal{O}(D^7)}$] it is still an interesting alternative way of defining the isometry condition. One advantage of YB-isoTNS is the ease with which the Ansatz can be generalized to different lattices. For instance, we expect the Ansatz to be extendable to the kagome lattice without increasing the cost scaling of the YB move. Moreover, more involved algorithms like DMRG require the contraction of a boundary MPS, which has a computational complexity scaling as ${\mathcal{O}(D^{10})}$ to ${\mathcal{O}(D^{12})}$ depending on the implementation. For such algorithms, we expect the increased cost for the YB move compared to the MM to not make a large difference. Furthermore, the algorithm for shifting the orthogonality hypersurface could be adapted to perform all YB moves in parallel by utilizing the canonical form of MPS introduced by Vidal~\cite{vidal_efficient_2003}. This is akin to the parallel implementation of TEBD on MPS~\cite{urbanek2016parallel, volokitin_propagating_2019, sun_improved_2024}. We anticipate that such a parallelized version would yield results equivalent to those of the sequential algorithm in the regime of low truncation error, while offering significant computational speedups for large system sizes. In contrast, a similar parallel strategy is not feasible for MM-isoTNS, highlighting a fundamental conceptual distinction between the two Ansätze: the YB move can be understood as entirely local, whereas the unzipping sweep in MM-isoTNS is a more global operation. \red{Consequently, we anticipate that extending YB-isoTNS to infinite strip geometries, as in Ref.~\cite{wu_two-dimensional_2023}, should be straightforward.} \par
It would be interesting to implement a DMRG algorithm on YB-isoTNS and compare its performance to TEBD. Future research could explore alternative algorithms for performing the YB move, which we find to be the main source of errors. Similar to MM-isoTNS, implementing a variational optimization on the complete column when shifting the orthogonality surface may be beneficial, as this was found to be essential for achieving good results when performing real time evolution with MM-isoTNS~\cite{lin_efficient_2022}. The algorithm could also significantly benefit from an implementation on Graphics Processing Units (GPUs), which can greatly accelerate tensor contractions and decompositions while also increasing the power efficiency. 
Additional promising research directions include the extension of the method to the thermodynamic limit and further investigation of the connection between isoTNS and quantum computing~\cite{liu_simulating_2024, slattery_quantum_2021}. Due to the direct correspondence between sequential quantum circuits and isoTNS~\cite{lin2021real} one can use the Ansatz to classically optimize sequential circuits. Furthermore, it would be interesting to study the limitations of isoTNS compared to unrestrained PEPS more systematically, for example using fermionic Gaussian states~\cite{wu_alternating_2025}.

\begin{acknowledgments}
The authors thank S.-H.~Lin, S.~Anand, and J.~Hauschild for helpful discussions. This work was supported by the Deutsche Forschungsgemeinschaft (DFG, German Research Foundation) under Germany’s Excellence Strategy EXC-2111-390814868, TRR360-492547816, and the Munich Quantum Valley, which is supported by the Bavarian state government with funds from the Hightech Agenda Bayern Plus. M.\,P. Zaletel was supported by the U.S. Department of Energy, Office of Science, Basic Energy Sciences, under Early Career Award No. DE-SC0022716.
\end{acknowledgments}

\textbf{Data and materials availability.} Data analysis and simulation codes are available on Zenodo~\cite{zenodo_YB_isoTNS}. The YB-isoTNS source code is openly available on github~\cite{github_YB_isoTNS}.

\bibliography{biblio}

\appendix
\section{Optimization of isometric tensor networks}
\label{app:optimization_of_isometric_tensor_networks}
When working with isometric tensor networks, one often needs to find optimal tensors that extremize a given cost function $f$. In the most general case, $f$ is a function
\begin{equation}
\label{eq:general_optimization_problem_of_isometric_tensor_networks_cost_function_multiple_input_tensors}
    f:\mathbb{C}^{m_1\times n_1}\times\mathbb{C}^{m_2\times n_2}\times \dots \times \mathbb{C}^{m_K\times n_K} \to \mathbb{R},
\end{equation}
mapping $K$ tensors ${T_1,\dots,T_K}$ to a scalar cost value. Here, the tensors have already been reshaped into matrices, grouping together legs with incoming arrows and legs with outgoing arrows, respectively. The tensors must satisfy certain constraints. If a tensor $T_i$ possesses both legs with incoming arrows and legs with outgoing arrows, it must satisfy the isometry constraint ${T_i^\dagger T_i = \id}$. If instead the tensor $T_j$ posesses only legs with incoming arrows (and thus is the orthogonality center), it is constrained to be normalized to one: ${\lVert T_j\rVert_\text{F} = 1}$. To summarize, we want to solve the optimization problem
\begin{equation}
	\label{eq:general_optimization_problem_of_isometric_tensor_networks_multiple_input_tensors}
	T_1^\text{opt}, \dots, T_K^\text{opt} = \underset{T_1,\dots,T_K}{\text{argmax}}f\left(T_1, \dots, T_K\right)
\end{equation}
under the constraints
\begin{equation}
	\label{eq:general_optimization_problem_of_isometric_tensor_networks_isometry_constraint}
	T_i^\dagger T_i = \id
\end{equation}
for isometries $T_i$ and
\begin{equation}
	\label{eq:general_optimization_problem_of_isometric_tensor_networks_ortho_center_constraint}
	\lVert T_j\rVert_\text{F} = 1
\end{equation}
for the orthogonality center $T_j$. \par
In the following, we will discuss several approaches for solving optimization problem \eqref{eq:general_optimization_problem_of_isometric_tensor_networks_multiple_input_tensors}. We will first assume that the input of the cost function is a single tensor $T$. If the cost function is linear, the problem is known as the \textit{orthogonal Procrustes problem}~\cite{gower2004procrustes} and we discuss its closed form solution in Sec.~\ref{app:Orthogonal_procrustes_problem}. Non-linear cost functions and functions of multiple tensors can be optimized by using the Evenbly-Vidal algorithm, see Sec.~\ref{app:EV_style_optimization}. \par
A different, more involved approach to solving the optimization problem is given by Riemannian optimization, which we discuss in Sec.~\ref{app:Riemannian_optimization}.

\subsection{Orthogonal Procrustes problem}
\label{app:Orthogonal_procrustes_problem}

If the cost function is linear, it can be written as
\begin{equation}
	f(T) = \sum_{j=1}^{m}\sum_{k=1}^{n}\left[\alpha_{j,k}\Re\left(T_{j,k}\right) + \beta_{j,k} \Im\left(T_{j,k}\right)\right]
\end{equation}
with parameters ${\alpha_{j,k}, \beta_{j,k} \in \mathbb{R}}$. Introducing the \textit{environment tensor} ${E\in\mathbb{C}^{m\times n}}$ as ${E_{j,k} = \alpha_j + i \beta_k}$, we can write the cost function as
\begin{equation}
	f(T) = \sum_{j=1}^{m}\sum_{k=1}^{n} \Re\left(E_{j,k}^*T_{j,k}\right) = \Re\Tr\left(E^\dagger T\right) = \Re\Tr\left(T^\dagger E\right).
\end{equation}
Maximizing ${f(T)}$ under the isometry constraint ${T^\dagger T = \id}$ is known as the orthogonal Procrustes problem and permits the closed form solution
\begin{equation}
	\label{eq:orthogonal_procrustes_problem_closed_form_solution}
	T_\text{opt} = \underset{T^\dagger T = \id}{\argmax} \Re\Tr\left(T^\dagger E\right) = UV^\dagger,
\end{equation}
where the matrices $U$ and $V$ are computed using an SVD ${E = USV^\dagger}$. To prove this result we insert the SVD into the cost function as
\begin{equation}
	\begin{split}
	f(T) &= \Re\Tr\left(ET^\dagger\right) = \Re\Tr\left(USV^\dagger T^\dagger\right) \\
        &= \Re\Tr\left[\left(U\sqrt{S}\right)\left(\sqrt{S}V^\dagger T^\dagger\right)\right] \\
	&= \Re\left\langle\sqrt{S}U^\dagger,\sqrt{S}V^\dagger T^\dagger\right\rangle_\text{F}.
	\end{split}
\end{equation}
We next use the fact that the Frobenius inner product satisfies the Cauchy-Schwarz inequality to obtain the upper bound
\begin{equation}
	\begin{split}
		f(T) &= \Re\left\langle\sqrt{S}U^\dagger,\sqrt{S}V^\dagger T^\dagger\right\rangle_\text{F} \le \left\lVert\sqrt{S}U^\dagger\right\rVert_\text{F}\left\lVert\sqrt{S}V^\dagger T^\dagger\right\rVert_\text{F} \\
		&= \sqrt{\Tr\left(USU^\dagger\right)\Tr\left(TVSV^\dagger T^\dagger\right)} = \Tr\left(S\right),
	\end{split}
\end{equation}
where in the last step we used ${U^\dagger U = \id}$, ${V^\dagger V = \id}$, ${T^\dagger T = \id}$, and the cyclic property of the trace. This upper bound is reached by the solution
\begin{equation}
	F\left(T_\text{opt}\right) = \Re\Tr\left(USV^\dagger VU^\dagger\right) = \Tr\left(S\right),
\end{equation}
proving \eqref{eq:orthogonal_procrustes_problem_closed_form_solution}. \par
If instead of satisfying the isometry condition the tensor $T$ must be normalized to one, the closed form solution can be found as
\begin{equation}
	\label{eq:orthogonal_procrustes_problem_simple_case_closed_form_solution}
	T_\text{opt} = \underset{\left\lVert T\right\rVert_\text{F} = 1}{\argmax} \Re\Tr\left(T^\dagger E\right) = E/\left\lVert E\right\rVert_\text{F}.
\end{equation}
We arrive at this solution through a similar argument as before. First, we obtain an upper bound
\begin{equation}
	f(T) = \Re\Tr\left(T^\dagger E\right) = \Re\left\langle T, E\right\rangle_\text{F} \le \left\lVert T\right\rVert_\text{F}\left\lVert E\right\rVert_\text{F} = \left\lVert E\right\rVert_\text{F}
\end{equation}
using the Cauchy-Schwarz inequality and the normalization constraint ${\left\lVert T\right\rVert = 1}$. We proceed by showing that the upper bound is reached by $T^\text{opt}$,
\begin{equation}
	f(T_\text{opt}) = \Re\Tr\left(EE^\dagger/\left\lVert E\right\rVert_\text{F}\right) = \left\lVert E \right\rVert_\text{F},
\end{equation}
proving \eqref{eq:orthogonal_procrustes_problem_simple_case_closed_form_solution}.

\subsection{Evenbly-Vidal style optimization}
\label{app:EV_style_optimization}

In general, the cost function ${f(T)}$ is not linear. For example, a nonlinear cost function is encountered in the disentangling procedure when optimizing a MERA wave function~\cite{evenbly_algorithms_2009}. It was proposed by Evenbly and Vidal~\cite{evenbly_algorithms_2009, evenbly_algorithms_2014} to linearize the cost function and to update the tensor $T$ iteratively using the closed form solutions from Sec.~\ref{app:Orthogonal_procrustes_problem}. Let us assume that the cost function ${f(T)}$ can be written as the contraction of a tensor network, where in general the tensor $T$ may appear multiple times. We contract all tensors except one of the tensors $T$ into an environment tensor ${E_T \in\mathbb{C}^{n\times n}}$ and the cost function becomes
\begin{equation}
	f(T)=\Re\Tr\left(E_TT\right).
\end{equation}
We now keep the environment $E_T$ fixed, treating it as if it were independent of $T$, and update $T$ with the closed form solutions \eqref{eq:orthogonal_procrustes_problem_closed_form_solution} or \eqref{eq:orthogonal_procrustes_problem_simple_case_closed_form_solution}. This is repeated until $T$ is converged. If and how fast $T$ converges depends on the details of the cost function, but convergence cannot be guaranteed for arbitrary cost functions. \par
Cost functions of multiple tensors ${T_1, \dots, T_K}$ can be optimized iteratively via an algorithm similar to the Evenbly-Vidal algorithm. Let us again assume that the cost function can be written as the contraction of a tensor network containing the tensors ${T_1, \dots, T_K}$. One can proceed by optimizing one tensor at a time as discussed above while keeping all other tensors fixed. If the cost function is linear in all tensors ${T_1, \dots, T_K}$ and bounded, ${f(T_1, \dots, T_K) \le c\in\mathbb{R}}$, this algorithm is guaranteed to converge, since each local update is optimal and thus the cost function can never decrease.\par

\subsection{Riemannian optimization}
\label{app:Riemannian_optimization}

\begin{figure}
    \includegraphics[scale=1]{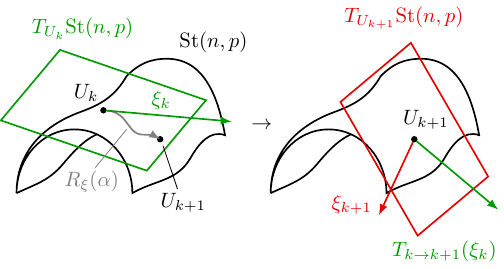}
    \caption{In this figure, a visualization of optimization on Riemannian manifolds is given. The iterate $X_k$ (left) is updated along the search direction $\xi_k$ , which is an element of the
    tangent space ${T_{X_k}\Stiefel}$. The next iterate $X_{K+1}$ is computed with the retraction ${R_{\xi_k}(\alpha)}$, where ${\alpha_k \in \mathbb{R}}$ is the step size. For the computation of the next search direction $\xi_{k+1}$ some algorithms need the previous search direction $\xi_{k}$, which must first be brought to the tangent space ${T_{X_{k+1}}\Stiefel}$ of the new iterate $X_{k+1}$ via the vector transport ${T_{k\rightarrow k+1}\left(\xi_k\right)}$.}
    \label{fig:Riemannian_optimization_sketch}
\end{figure}
In the following, we give a brief review of Riemannian optimization algorithms over the manifold of isometric matrices ${T\in\mathbb{C}^{n\times p}}$, ${T^\dagger T = \id}$, which is called the \textit{Stiefel manifold} $\Stiefel$. For a more in-depth introduction to the topic we recommend the book~\cite{absil_optimization_2008}. Riemannian optimization of complex matrix manifolds is discussed in the context of quantum physics and isometric tensor networks in Refs.~\cite{luchnikov_riemannian_2021, hauru_riemannian_2021}. An implementation of Riemannian optimization on the real Stiefel manifold and other matrix manifolds in python is given in~\cite{townsend_pymanopt_2016}. Some parts of this implementation were also used in our implementation~\cite{github_YB_isoTNS}. \par
A typical optimization algorithm iteratively improves an iterate ${X_k\in\Stiefel}$, ${k=1,2,\dots}$ until a local minimum of a cost function $f:\Stiefel\to\mathbb{R}$ is found. In Riemannian optimization, the gradient ${\bm{\nabla} f\left(X_k\right)}$ of the cost function is restricted to the tangent space ${T_{X_k}\Stiefel}$ of the iterate $X_k$~\cite{absil_optimization_2008}, which we visualize in Fig.~\figref{fig:Riemannian_optimization_sketch}. The gradient can be computed either analytically or via automatic differentiation~\cite{luchnikov_riemannian_2021, townsend_pymanopt_2016}. Optimization algorithms typically compute a search direction ${\xi_k\in T_{X_k}\Stiefel}$ and a step size ${\alpha_k \in \mathbb{R}}$ from the gradient. In an optimization algorithm defined on a Euclidean vector space one would then move along this direction as
\begin{equation}
	\tilde{X}_{k+1} = X_k + \alpha_k\xi_k.
\end{equation}
However, $\tilde{X}_{k+1}$ is in general not an element of the manifold. To ensure ${X_{k+1}\in\Stiefel}$, one can introduce a \textit{retraction} ${R_{\xi_k}:\mathbb{R}\to\Stiefel}$~\cite{absil_optimization_2008}. One can think of ${R_{\xi_k}(\alpha_k)}$ as moving along the direction of $\xi_k$ while staying on the manifold. As $\alpha_k$ increases, we move further along the path defined by the retraction, with ${R_{\xi_k}(0) = X_k}$; see Fig.~\figref{fig:Riemannian_optimization_sketch}. Different retractions can be chosen, varying by how well they perform in optimization problems and by how hard they are to compute. A retraction for the Stiefel manifold that is particularly easy to compute while still yielding good results in practice is given by
\begin{equation}
	R_{\xi_k}(\alpha_k) = \qf\left(X_k + \alpha_k\xi_k\right),
\end{equation}
where ${\qf\left(A\right)}$ is the $Q$ factor of the QR decomposition ${A = QR}$.\par
Many optimization algorithms, e.g. conjugate gradients, require gradients or search directions from previous iterates for computing a search direction at the current iterate. In Riemannian optimization, one must first bring these gradients and search directions from the tangent spaces of previous iterates to the tangent space of the current iterate. This is handled by a so-called \textit{vector transport} ${T_{k\rightarrow k+1}\left(\xi_k\right)}$~\cite{absil_optimization_2008}; see Fig.~\figref{fig:Riemannian_optimization_sketch}. \par
Finally, for optimization algorithms of second order such as the trust-region method, one needs to generalize the notion of the Hessian-vector product to Riemannian manifolds. This generalization is given by the \textit{Riemannian connection}~\cite{absil_optimization_2008}. For the Stiefel manifold the Hessian-vector product is simply given by projecting the Hessian-vector product of the embedding Euclidean vector space ${\mathbb{C}^{n\times p}}$ to the tangent space~\cite{absil_optimization_2008}. \par
In this work we use two algorithms for performing Riemannian optimization: conjugate gradients (CG) and the trust-region method (TRM); see also our implementation at Ref.~\cite{github_YB_isoTNS}.\par

\subsubsection{Conjugate gradients}
CG uses the accumulated gradients of previous iterations to compute an improved search direction, trying to achieve faster convergence than simple gradient descent~\cite{shewchuk1994introduction, hager2006survey, hager_algorithm_2006, absil_optimization_2008}. For an in-depth explanation of how CG on the Stiefel manifold can be implemented, see Refs.~\cite{absil_optimization_2008, zhu_riemannian_2017, hauru_riemannian_2021}. In our implementation we additionally use Powell's restart strategy~\cite{kou2014use, hager2006survey, townsend_pymanopt_2016}, which can significantly improve the efficiency of CG. \par

\subsubsection{Trust region methods}
Trust-region methods are a class of second order optimization techniques that are known for their desirable global convergence properties with a local superlinear rate of convergence~\cite{absil_optimization_2008, absil_trust-region_2007}. The main idea of TRM is to locally approximate the cost function $f$ by a quadratic model ${m_{X_k}(\eta) \approx f(X_k+\eta)}$ with ${\eta\in T_{X_k}\Stiefel}$ in a \textit{trust region} ${\langle\eta,\eta\rangle \le \Delta_k^2}$ of radius $\Delta_k$ around the current iterate $X_k$. To define this model, one must compute the Hessian-vector product or an approximation thereof. This approximate cost function is then minimized within the trust region using truncated conjugate gradients (tCG), which converges quickly for the quadratic approximation. Depending on the quality of the approximation at the current iterate one can shrink or enlarge the trust region. TRM is able to achieve local superlinear convergence while still retaining global convergence properties~\cite{absil_optimization_2008} and can be thought of as an improved version of Newton's method. For more details on the TRM on Riemannian manifolds, see Refs.~\cite{absil_optimization_2008, absil_trust-region_2007, townsend_pymanopt_2016}.
\end{document}